\documentstyle[12pt]{article}

\input epsf

\begin{document}

\def\stacksymbols #1#2#3#4{\def\theguybelow{#2}
    \def\verticalposition{\lower#3pt}
    \def\spacingwithinsymbol{\baselineskip0pt\lineskip#4pt}
    \mathrel{\mathpalette\intermediary#1}}
\def\intermediary#1#2{\verticalposition\vbox{\spacingwithinsymbol
      \everycr={}\tabskip0pt
      \halign{$\mathsurround0pt#1\hfil##\hfil$\crcr#2\crcr
               \theguybelow\crcr}}}

\def\lapproxeq{\stacksymbols{<}{\sim}{2.5}{.2}}
\def\gapproxeq{\stacksymbols{>}{\sim}{3}{.5}}

\begin{titlepage}

\begin{flushright}
MIT-CTP-2449\\
TUTP-95-2\\
hep-th/9507035\\
June 1995
\end{flushright}

\vskip 0.9truecm

\begin{center}
{\large {\bf Quantum gravity effects at a black hole horizon$^*$}}
\end{center}

\vskip 0.8cm

\begin{center}
{Gilad Lifschytz}
\vskip 0.2cm
{\it Center for Theoretical Physics,
    Laboratory for Nuclear Science\\
    and Department of Physics,\\
    Massachusetts Institute of Technology,
    Cambridge MA 02139, USA.\\
    e-mail: gil1@mitlns.mit.edu }
\vskip 0.3cm
{and}
\vskip 0.3cm
{Miguel Ortiz}
\vskip 0.2cm
{\it Institute of Cosmology,
Department of Physics and Astronomy,\\
Tufts University,
Medford, MA 02155, USA.\\
e-mail: mortiz@cosmos2.phy.tufts.edu }
\end{center}

\vskip 0.7cm

\noindent {\small {\bf Abstract:} Quantum fluctuations in the background
geometry of a black hole are shown to affect the propagation of matter
states falling into the black hole in a foliation that corresponds to
observations purely outside the horizon. A state that starts as a Minkowski
vacuum at past null infinity gets entangled with the gravity sector, so
that close to the horizon it can be represented by a statistical ensemble
of orthogonal states.  We construct an operator connecting the different
states and comment on the possible physical meaning of the above
construction. The induced energy-momentum tensor of these states is
computed in the neighbourhood of the horizon, and it is found that
energy-momentum fluctuations become large in the region where the bulk of
the Hawking radiation is produced. The background spacetime as seen by an
outside observer may be drastically altered in this region, and an outside
observer should see significant interactions between the infalling matter
and the outgoing Hawking radiation. The boundary of the region of strong
quantum gravitational effects is given by a time-like hypersurface of
constant Schwarzschild radius $r$ one Planck unit away from the horizon.
This boundary hypersurface is an example of a stretched horizon.}

\rm
\noindent
\vskip 0.7 cm

\begin{flushleft}
$^*$ {\small This work was supported in part by funds
provided by the U.S. Department of Energy (D.O.E.) under
cooperative agreement DE-FC02-94ER40818, and in part by the
National Science Foundation.}
\end{flushleft}

\end{titlepage}

\section{Introduction}

The standard derivation of the Hawking radiation of a black hole involves the
propagation of free fields on an idealised collapsing black hole
background \cite{hawking}. This simple calculation indicates that
black holes evaporate through approximately thermal radiation. Unless there
are subtle correlations between the matter that forms the black hole and
the outgoing radiation that are ignored in this calculation, there is a net
loss of information after the evaporation is complete.

Correlations between the degrees of freedom that form the black hole and
those of the Hawking radiation appear to be ruled out by simple arguments
\cite{preskill}. According to the equivalence principle, matter freely
falling into the black hole should not be subject to any extraordinary
interactions as it passes through the event horizon. Thus the information
carried by the infalling matter is carried into the black hole interior,
and becomes inaccessible to an outside observer. Unless the information can
be simultaneously contained in parts of the quantum state inside and
outside the black hole, it is lost behind the event horizon.

Ever since Hawking's original derivation, there have been attempts to show
that quantum gravity should affect the evaporation process in such a way
that information is in fact conserved \cite{page}. The preceding arguments
indicate that if quantum gravity is to play a role in resolving the black
hole information paradox, its effects must become important at the event
horizon. However, the outgoing Hawking state is in the Unruh or Kruskal
vacuum close to the horizon, and quantum field theory calculations have
clearly shown that there are no large curvature or back-reaction effects
there \cite{BnD}. The only exception comes at the very late stages of
evaporation, when the black hole becomes Planck-sized, and the receding
horizon approaches the strong curvature region within the black hole
core. However, at this late stage, it is very difficult to arrange for
quantum gravity effects to restore information within a finite time
\cite{casher}.

Over the last few years, a number of new suggestions have been made of how
non-perturbative quantum gravitational effects might be responsible for new
physics at the black hole horizon. The most ardent proponent of this
viewpoint has been 't Hooft \cite{thooft} who as early as 1985 suggested
that interactions between infalling matter and the outgoing Hawking
radiation should not be ignored, since they involve Planckian energy
scattering processes which can only be computed in a quantum theory of
gravity. More recently several authors have added pieces to the puzzle, and
a coherent picture of how quantum gravity could be important close to a
black hole horizon has begun to emerge [6-13].

An important conceptual step was taken by 't Hooft and Susskind in
introducing the notion of black hole complementarity. Susskind {\it
et. al.} \cite{stu} point out that large quantum gravitational effects
close to the horizon are only consistent if coordinate transformations
involving boosts on the oder of the Planck scale, transform the matter
states in an unconventional way. It is conjectured that in this way,
freely-falling observers could experience no large effects near the
horizon, while outside observers at rest with respect to the horizon
perceive it as a region of strong interaction where semiclassical physics
breaks down and infalling states interact significantly with the outgoing
radiation.

A series of calculational frameworks have emerged which allow the quantum
gravitational effects predicted by 't Hooft to be computed and which
indicate the possibility of corrections to the standard Hawking
derivation. It was first noted by 't Hooft \cite{thooft} that the effect of
planckian scattering between infalling and outgoing matter was to cause a
shift in Kruskal coordinates $x^-\to a^-(x^- + \Delta^-)$ in the outgoing
Hawking modes. A large shift of this kind has a noticeable effect on the
outgoing radiation, and indeed in 2-d models \cite{cghs} the shift is
responsible for the Hawking radiation in the usual semi-classical
formalism. However, when considering the consequences of small fluctuations
in the infalling matter, the shift is unimportant since Hawking radiation
is approximately in the Kruskal vacuum close to the horizon. The
transformation $x^-\to a^-(x^- + \Delta^-)$ is thus an approximate symmetry
of the Hawking state and the only effect on the radiation as observed at
infinity is a small shift in its temperature \cite{ver1}.

More recently, in Refs. \cite{samir,klmo} it was pointed out that large
quantum gravitational effects are observed in the propagation of matter
fields falling into the horizon. In Ref. \cite{klmo} it was shown that an
inherent Planck scale uncertainty in the mass of a black hole (due to
quantum fluctuations in the infalling matter that forms the black hole)
leads to a large uncertainty in the state of matter falling through the
horizon on a particular set of spacelike hypersurfaces which we shall refer
to as S-surfaces. These are surfaces that capture some of the Hawking
radiation at future null infinity, avoid the last stages of evaporation of
the hole and then stay outside the horizon in the region above the
shock-wave (see Fig. 1). An important element in this calculation is a
separation between the matter that forms the hole and that which falls
through the horizon after the hole is formed. It is found that a small
fluctuation in the black hole mass leads to a shift in both the in- and
out-going modes. In a region of the S-surface close to the horizon but not
too close to the past horizon, the change in the modes can be expressed in
Kruskal coordinates as $x^-\to a^-(x^- + \Delta^-)$ and $x^+\to a^+(x^+
+\Delta^+)$, where $\Delta^\pm$ are proportional to the shift in the mass,
but are also highly sensitive to the properties of the S-surface.  The
shift in $x^-$ is small for small fluctuations in mass, so that by the
arguments given above there is a negligible effect on the outgoing radiation
in this first approximation. However, since the infalling matter is
typically found in a state close to the Schwarzschild vacuum, the
transformation in $x^+$ has a non-trivial effect.  For a fluctuation as
small as $\Delta M\sim e^{-M}$ in Planck units, the shift in $x^+$ and the
size of the region for which it is valid are large enough to change the
incoming state appreciably.

In Sec. 2.1 we present a review of the derivation of the shifts $x^-\to
a^-(x^- + \Delta^-)$ and $x^+\to a^+(x^+ +\Delta^+)$ whose origin is the
pointwise identification of the same spacelike hypersurface in two
different classical spacetimes according to its intrinsic geometry. In
Sec. 2.1 the shift is derived for CGHS \cite{cghs} solutions with different
masses. This is generalised in Sec. 2.2 to the identification of
spherically symmetric spacelike hypersurfaces in Schwarzschild spacetimes
with different masses.

In Sec. 3 the effect of the shift on the propagation of matter is
discussed.  Although in this paper we will talk mainly about the CGHS
model, the results given in Sec. 2.2 show that similar results hold for
matter in the S-wave on a Schwarzschild background. The range of quantum
fluctuations in the mass of a black hole is determined by the finite
lifetime and size of the hole, implying an uncertainty in the mass $M$ of
at least $1/M$.  These fluctuations are extremely small and should not be
observable, so we can consider the effect of tracing over them. The result
is a density matrix of the form
\begin{equation}
\rho(\Sigma)=\int dM |\omega(M)|^2
|f_M(\Sigma)\rangle
\langle f_M(\Sigma)|
\end{equation}
where $\omega(M)$ is some weight function peaked around $M_0$ with a width
of order 1, and each $|f_M(\Sigma)\rangle$ is obtained by propagating the
matter state from the vacuum state at past null infinity to the S-surface
$\Sigma$ on a black hole background of mass $M$.

A semiclassical interpretation of this density matrix requires that one
define a mean background spacetime (say that with mass $M_0$)\footnote{Note
that the choice of mean mass $M_0$ is somewhat arbitrary since we could
choose any mass $M$ within the peak of the weight function $\omega(M)$}, on
which $\rho$ evolves, and that one regard each $|f_M(\Sigma)\rangle$ with
$M\ne M_0$ as an element of the Hilbert space ${\cal H}_{{\cal
M}_0,\Sigma}$ of states on $\Sigma$ in the spacetime ${\cal M}_0$ with mass
$M_0$. This can be achieved by using the canonical inner product between
states defined on the same spacelike hypersurface embedded in different
classical spacetimes \cite{klmo}.

\medskip
\begin{center}
\leavevmode
\epsfxsize 2in
\epsfbox{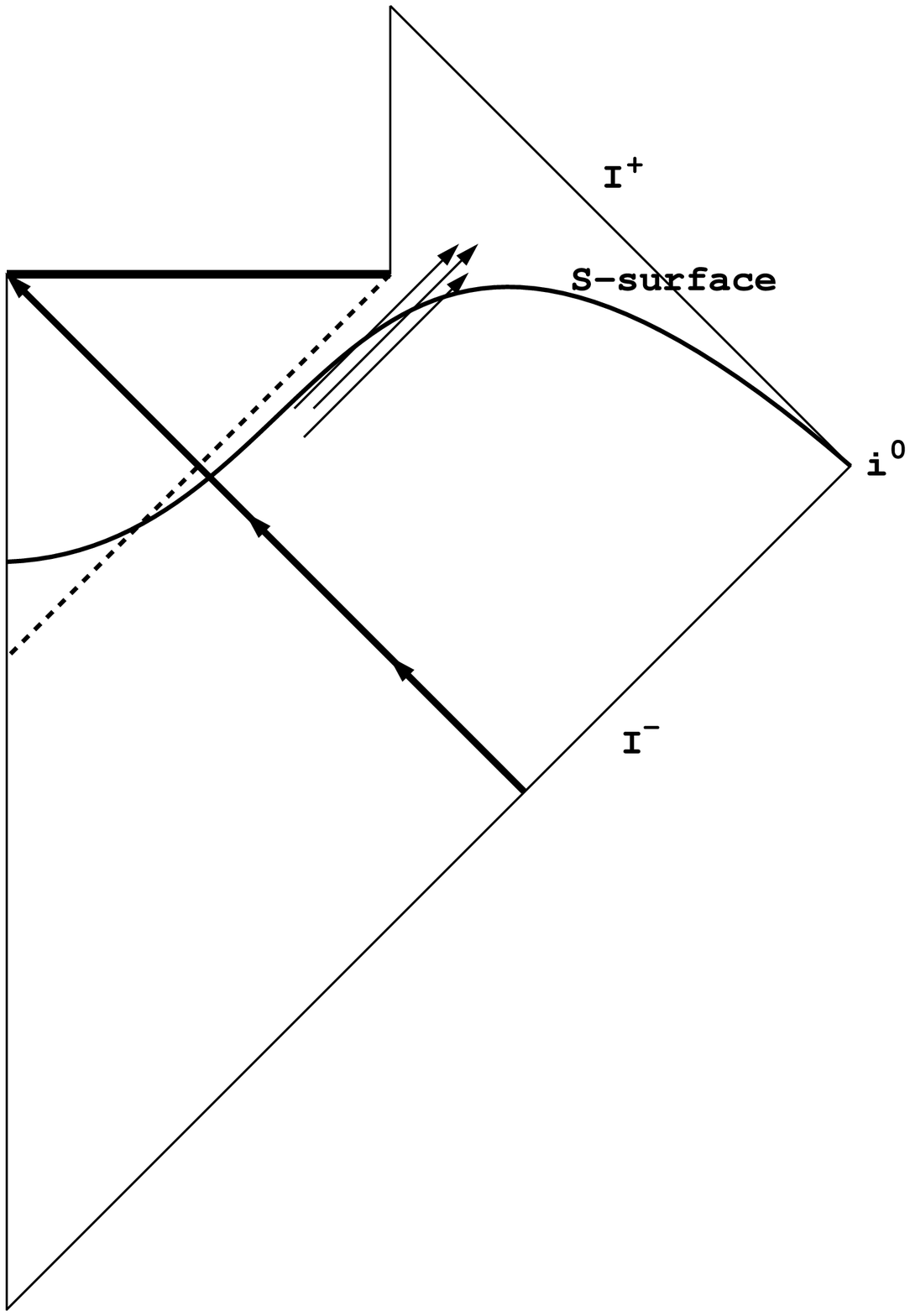}
\end{center}
\vskip 0.5 true cm
\begin{quotation}
\small
\noindent {\bf Figure 1:} An S-surface shown in an evaporating black hole
spacetime. The surface remains outside the event horizon in the region
above the shock-wave. The arrows represent the outgoing Hawking radiation.
\end{quotation}
\medskip

The semiclassical approximation as it is generally used, assumes that all
the $|f_M(\Sigma)\rangle$ are approximately equal. In that case $\rho$ is a
pure state density matrix that evolves according to the functional
Schr\"odinger equation on any background spacetime with a mass close to the
mean black hole mass $M_0$. This is generally true for generic
hypersurfaces $\Sigma$. However, on a set of S-surfaces the shift $x^+ \to
a^+(x^+ +\Delta^+)$ causes a change in the $|f_M(\Sigma)\rangle$ that is
very sensitive to the value of $M$. Then $\rho$ is a mixed state reflecting
the entanglement of the matter with the quantum gravitational
fluctuations. The change in the $|f_M(\Sigma)\rangle$ is concentrated in an
extended region of the S-surface adjacent to the horizon, and is large
enough that a pair of states $|f_M(\Sigma)\rangle$ and
$|f_{\bar{M}}(\Sigma)\rangle$ are approximately orthogonal if
$M-\bar{M}\gapproxeq e^{-M}$ \cite{klmo}. This can be taken to indicate a
breakdown in the semiclassical approximation.

In Sec. 3 it is shown how each element $|f_M(\Sigma)\rangle$ defines a Fock
state on ${\cal M}_0$:
\begin{equation}
{\cal U}(\Sigma,M\to M_0)|f_{M_0}\rangle
\end{equation}
which is defined to be equal to $|f_M(\Sigma)\rangle$ when evaluated on
$\Sigma$ in ${\cal M}_0$. Here $|f_{M_0}\rangle$ is a Fock state that
agrees with $|f_M({\cal I}^-)\rangle$ when evaluated at past null
infinity. The operator ${\cal U}$ implements the shift $x^+ \to a^+(x^+
+\Delta^+)$ and is constructed explicitly in Sec. 4. It is shown to bear a
close resemblance to the operator defined in Ref. \cite{kvv} to describe
the Planckian interactions between in and outgoing matter at the horizon.

The inner product alone does not give any sense of how the shift $x^+\to
a^+(x^++\Delta^+)$ changes the picture of black hole evaporation, since
orthogonal states can give rise to very similar physics. In Sec. 5 we
present a simple calculation of the energy associated with the orthogonal
infalling states, which we shall use to argue both that the breakdown in
the semiclassical approximation is large, and that a backreaction on the
outgoing Hawking radiation is produced. We shall compute the energy flux of
the orthogonal states on an entire S-surface including the region close to
the infalling matter that formed the black hole.  The energy flux is zero
below the shock wave and far above it, but in between it becomes large
enough to affect the outgoing Hawking radiation at its early stages.

In Sec. 6 we define a region, determined by the S-surfaces and by the
location within the S-surfaces of the large energy fluctuations, in which
there is an irrevocable breakdown in the semiclassical approximation. This
region is shown to be bounded by a surface approximately one Planck
distance from the horizon, resembling a stretched horizon. This stretched
horizon represents a boundary beyond which we cannot safely perform
calculations armed only with a knowledge of semiclassical physics. An
important aspect of our approach, however, is that this breakdown is only
manifested when working in a very specific foliation close to the horizon,
and a different choice of foliation allows a semiclassical description of
horizon physics. We interpret this as a manifestation of black hole
complementarity \cite{stu}, in the sense that different observers have
different perceptions of the effects of spacetime fluctuations. For
observers that stay outside the horizon and detect some of the Hawking
radiation, quantum gravitational effects are large, and there is a strong
interaction between infalling matter and the outgoing radiation.

\section{A Large Shift}

\subsection{The CGHS black hole}

In this subsection we briefly review the geometrical origin of the large
shift in Kruskal coordinates \cite{klmo} for the CGHS black hole
\cite{cghs}. This shift comes from the identification, through its
intrinsic geometry, of a single spacelike hypersurface $\Sigma$ embedded in
different classical spacetimes. The main result we rederive in this
subsection is given in Eqs. (\ref{lshifts}) and (\ref{gshifts}).

We shall work in the CGHS spacetime \cite{cghs},
\begin{eqnarray}
&&ds^2= -e^{2\rho}dx^+ dx^-,
\nonumber
\\
&&e^{-2\rho}=e^{-2\phi}= -{M\over\lambda}(\lambda x^+-1)
\Theta(x^+-1/\lambda)-\lambda^2 x^+ x^-
\label{cghs}
\end{eqnarray}
where $\phi$ is a dilaton field that can be taken to represent the radial
component of a four-dimensional metric after dimensional reduction
\cite{harstro}. Eq. (\ref{cghs}) represents a black hole formed by the collapse
of a shock wave of matter.  The $x^{\pm}$ are referred to as Kruskal
coordinates.

The intrinsic geometry of a spacelike hypersurface in the CGHS model can be
defined by the function $\phi(s)$ where $s$ is the proper distance to some
fixed point on the hypersurface determined by $\rho$. The fixed point
should be defined in a canonical fashion. It is natural to consider te
fixed point to be at infinity, since this corresponds to anchoring the
surface at infinity, thereby ensuring the validity of the semiclassical
approximation there. Two hypersurfaces embedded in different spacetimes
have the same intrinsic geometry if they define the same function
$\phi(s)$.

Consider two different solutions (\ref{cghs}) with masses $M$ and
$\bar{M}$.  We can define a spacelike hypersurface $\Sigma$ in the first
spacetime through a relation between $x^+$ and $x^-$, say $\lambda
x^-=g(\lambda x^+)-M/\lambda$ (where this definition is made for
convenience as the horizon is at $\lambda x^-=-M/\lambda$).  Given
(\ref{cghs}), it is easy to deduce the function $\phi(s)$ which defines its
intrinsic geometry. Consider also a hypersurface $\bar{\Sigma}$ in the mass
$\bar{M}$ spacetime, defined by the relation $\lambda
\bar{x}^-=\bar{g}(\lambda \bar{x}^+)-\bar{M}/\lambda$. When $\Sigma$ and
$\bar{\Sigma}$ define the same $\phi(s)$ in two different spacetimes, there
is a canonical map between points on the hypersurface in each of the
spacetimes which follows from identifying local intrinsic geometry. The
condition that $\phi(s)$ is the same for $\Sigma$ and $\bar{\Sigma}$ is
equivalent to the two conditions:
\begin{eqnarray}
&& \frac{M}{\lambda} - \lambda x^{+}g(\lambda x^{+})=
 \frac{\bar{M}}{\lambda} - \lambda \bar{x}^{+}\bar{g} (\lambda
 \bar{x}^{+})
\label{g1}
\\ && \frac{g(\lambda x^{+}) + \lambda x^{+} g'(\lambda x^{+})
}{\sqrt{-g'(\lambda x^{+})}} = \frac{\bar{g}(\lambda \bar{x}^{+}) +
\lambda \bar{x}^{+} \bar{g}'(\lambda \bar{x}^{+})
}{\sqrt{-\bar{g}'(\lambda \bar{x}^{+})}}
\label{g2}
\end{eqnarray}
This set of equations can be solved for $\bar{g}(\lambda\bar{x}^{+})$ given
$g(\lambda x^+)$. The identification of points with the same local
intrinsic geometry is given by functions $\bar{x}^{+}(x^{+})$ and
$\bar{x}^{-}(x^{-})$. These functions are essential for comparing states
evolved to $\Sigma$ and $\bar{\Sigma}$ in the two different spacetimes.

The particular class of surfaces that we focus on in this paper, referred
to in the introduction as S-surfaces, can be specified by
\begin{equation}
g(\lambda x^+)=-\alpha^{2}(\lambda x^{+}-1) -\delta
\label{small}
\end{equation}
The two constants, the slope $\alpha$, and the distance $\delta$ in $x^-$
coordinates between the horizon (at $\lambda x^-=-M/\lambda$) and the point
where the surface crosses the shock wave (at $\lambda x^+=1$), are chosen
so that $\alpha^{2}\ll 1$ and $\delta\ll 1$, ensuring that the surface is
almost lightlike and runs very close to the horizon.

A particularly simple choice of $g(\lambda x^+)$ is
\begin{equation}
\begin{array}{rcll}
\displaystyle \lambda x^- &=& \displaystyle -\alpha^2\lambda x^+ -2\alpha
\sqrt{M\over\lambda} -{M\over \lambda}\qquad
&\displaystyle \lambda x^+\ge 1
\\
\displaystyle \lambda x^-&=& \displaystyle
-\left(\alpha+\sqrt{M\over\lambda}\right)^2\lambda{x}^+
&\lambda x^+\le 1
\end{array}
\label{line}
\end{equation}
For this choice of surfaces the scale of $\alpha$ is fixed by the condition
that the S-surface capture a proportion $r$ of the Hawking radiation, which
requires that $\alpha\sim e^{-r M/\lambda}$.

For (\ref{line}) we find the following relations $\bar{x}^+(x^+)$. For
$\bar{M}<M$:
\begin{equation}
\begin{array}{rcll}
\displaystyle \lambda \bar{x}^+ &=& \displaystyle \lambda x^+ +
{1\over\alpha}\left(\sqrt{M\over\lambda}
-\sqrt{\bar{M}\over\lambda}\right)
&\displaystyle \lambda x^+\ge 1
\\
\displaystyle \lambda \bar{x}^+ &=& \displaystyle
\left(1+ {1\over\alpha}\sqrt{M\over\lambda}\right)
\lambda x^+ -{1\over\alpha}\sqrt{\bar{M}\over\lambda}\qquad
&\displaystyle 1\ge\lambda x^+\ge \lambda x^+_<
\\
\displaystyle \lambda \bar{x}^+ &=&
\displaystyle \left({\alpha+\sqrt{M/\lambda}\over
\alpha+\sqrt{\bar{M}/\lambda}}\right) \lambda x^+
&\displaystyle \lambda x^+_<\ge\lambda x^+\ge 0
\end{array}
\label{lshifts}
\end{equation}
where $\lambda x^+_<=(\alpha
+{\scriptstyle\sqrt{\bar{M}/\lambda}})/(\alpha+
{\scriptstyle\sqrt{M/\lambda}})$. For $\bar{M}>M$:
\begin{equation}
\begin{array}{rcll}
\displaystyle
\lambda \bar{x}^+ &=& \displaystyle \lambda x^+ -
{1\over\alpha}\left(\sqrt{\bar{M}\over\lambda}
-\sqrt{M\over\lambda}\right)
&\displaystyle \lambda x^+\ge \lambda x^+_>
\\
\displaystyle \lambda \bar{x}^+ &=& \displaystyle
\left(1+ {1\over\alpha}\sqrt{\bar{M}
\over\lambda}\right)^{-1}
\left(\lambda x^+ +{1\over\alpha}\sqrt{M\over\lambda}\right)\qquad
&\displaystyle\lambda x^+_>\ge\lambda
x^+\ge 1
\\
\displaystyle \lambda \bar{x}^+ &=& \displaystyle
\left({\alpha+\sqrt{M/\lambda}\over
\alpha+\sqrt{\bar{M}/\lambda}}\right) \lambda x^+
&\displaystyle 1\ge\lambda x^+\ge 0
\end{array}
\label{gshifts}
\end{equation}
where $\lambda x^+_> = 1+
({\scriptstyle\sqrt{\bar{M}/\lambda}-\sqrt{M/\lambda}})/\alpha$.  The above
coordinate relationship translates directly to a relationship between
coordinates $\lambda v=\ln(\lambda x^+)$ and $\lambda\bar{v} =\ln(\lambda
\bar{x}^+)$ that are asymptotically flat on ${\cal{I}}^{+}$.  The relations
$\bar{x}^-(x^-)$ can be immediately derived from Eqs. (\ref{lshifts}) and
(\ref{gshifts}) given expressions for $\lambda x^-=g(\lambda
x^+)-M/\lambda$ and $\lambda \bar{x}^-=\bar{g}(\lambda
\bar{x}^+)-\bar{M}/\lambda$.

It should be apparent from (\ref{lshifts}) and (\ref{gshifts}) that a large
effect comes in each case from the first relation, for which
$\bar{x}^+\approx x^++\Delta^+$ with $\Delta^+$ of order $1/\alpha$. This
relation is valid for the region of the S-surface above the shock wave, all
the way out to $i_0$. A large effect also comes from the second
relation. Although we might be suspicious of the details of the metric in
the neighbourhood of the shock wave of the CGHS solution, and so mistrust
this second relation, it is inevitable that the relation involve large
parameters to compensate for the shift in the first relation, since the
third relation is in each case approximately the identity.

\subsection{The Schwarzschild black hole}

In this subsection we derive the large shift in Kruskal coordinates for the
case of a Schwarzschild black hole following the derivation given in the
previous subsection for the CGHS solution. The main result of this
subsection is given in Eq. (\ref{asycore}).

Taking $G=c=1$, the Schwarzschild metric can be written as
\begin{equation}
ds^{2}=-\frac{2M}{r}e^{-r/2M}dx^+dx^-+r^{2}d\Omega ^{2}
\label{metric}
\end{equation}
Here $\kappa x^+=e^{\kappa v}$ and $\kappa x^-=-e^{-\kappa u}$ where
$\kappa=1/4M$ is the surface gravity, $u=t-r^{*}$, $v=t+r^{*}$,
$r^{*}=r+2M\ln|r/2M-1|$ is the tortoise coordinate and $r$ is the usual
Schwarzschild coordinate. We shall refer to $x^+$ and $x^-$ as Kruskal
coordinates. It follows that
\begin{equation}
-\kappa x^+\kappa x^-=(r/2M-1)e^{r/2M}.
\end{equation}
For the purposes of this paper, we are only interested in the metric near
the future horizon, so we expand (\ref{metric}) around $r=2M$ by writing
$r=2M(1+{R})$, where $0<R\lapproxeq M_{pl}/M$, so that $\kappa x^+\kappa
x^-=-e{R}$.

If we write
\begin{equation}
ds^{2}=-e^{2\rho}dx^{+}dx^{-}+ Ad\Omega ^{2}
\end{equation}
then it follows that near the future horizon ({\it i.e.} when $-\kappa
x^+\kappa x^-\lapproxeq eM_{pl}/M$ and $\kappa x^+>1$),
\begin{equation}
e^{-2\rho}=e-2\kappa x^+\kappa x^-,\qquad
A={4M^{2}\over e}\left(e-2\kappa x^{+} \kappa x^{-}\right)
\label{met}
\end{equation}
This is similar to the CGHS metric in the coordinates of (\ref{cghs}) since
the dilaton field $e^{-2\phi}$ is the coordinate scalar that represents the
area of 2-spheres $A$ after dimensional reduction.  It is convenient from
now on to set $\hbar=1$ and work with a dimensionless $M$ defined in Planck
units so that $M\gg1$.

Let us now consider the matching of spacelike hypersurfaces in
Schwarzschild solutions with different masses. As a simplification, we
shall restrict observations of the matter fields to spherically symmetric
spacelike hypersurfaces. Such hypersurfaces are defined by their intrinsic
geometry in the $t,r$ plane.  The area $A$ of a 2-surface of constant $r$
is a scalar under a coordinate transformation in the $t,r$ plane. The
intrinsic geometry of a spherically symmetric hypersurface is thus given by
the function $A(s)$ where $s$ is the proper distance from infinity in the
$t,r$ plane, determined by $\rho$. Two hypersurfaces embedded in different
spacetimes have the same intrinsic geometry if they define the same
function $A(s)$.

As before, we now consider two black holes with masses $M$ and $\bar{M}$.
We can define a spherically symmetric spacelike hypersurface $\Sigma$ in
the first spacetime through a relation between $x^+$ and $x^-$, say
$x^-=g(\kappa x^+)$ (note that in (\ref{metric}) the horizon is at
$x^-=0$).  Given the form of the metric, it is easy to deduce the function
$A(s)$ which defines the intrinsic geometry. We then seek a hypersurface
$\bar{\Sigma}$ in the mass $\bar{M}$ spacetime, defined by the relation
$\bar{x}^-=\bar{g}(\bar{\kappa} \bar{x}^+)$ defining the same function
$A(s)$. This is equivalent to the two conditions:
\begin{equation}
A({\kappa x^+},g({\kappa
x^+}))=\bar{A}({\bar{\kappa}\bar{x}^+},\bar{g}({\bar{\kappa}\bar{x}^+}))
\label{c1}
\end{equation}
\begin{equation}
{dA\over ds}({\kappa x^+},g({\kappa x^+}))={d\bar{A}\over
d\bar{s}}({\bar{\kappa}\bar{x}^+},\bar{g}({\bar{\kappa}\bar{x}^+}))
\label{c2}
\end{equation}

Plugging (\ref{met}) into (\ref{c1}) and (\ref{c2}), one gets
\begin{eqnarray}
M^2\left(e-2\kappa x^{+} \kappa g\right)&=&
\bar{M}^2\left(e-2\bar{\kappa} \bar{x}^{+}\bar{\kappa}\bar{g}\right).
\label{line1} \\
\frac{\kappa g+\kappa x^{+}\kappa g'}{\sqrt{-\kappa g'}} & = &
\frac{\bar{\kappa}\bar{g}+\bar{\kappa} \bar{x}^{+} \bar{\kappa}\bar{g}'}
{\sqrt{-\bar{\kappa}\bar{g}'}}.
\label{line2}
\end{eqnarray}
Here a prime denotes differentiation with respect to $\kappa x^+$ or
$\bar{\kappa} \bar{x}^+$.  Differentiating equation (\ref{line1}) with
respect to $\kappa x^{+}$ and comparing with equation (\ref{line2}) we find
\begin{equation}
\frac{d \left(\bar{\kappa}\bar{x}^{+}\right)}{d \left(\kappa
x^{+}\right)}=\frac{M^2\sqrt{-\kappa
g'}}{\bar{M}^2\sqrt{-\bar{\kappa}\bar{g}'}}.
\label{diff}
\end{equation}
Equation (\ref{diff}) can be integrated using equations (\ref{line1}) and
(\ref{line2}) to give :
\begin{equation}
\ln (\bar{\kappa} \bar{x}^{+} )= {2M^2\over\bar{M}^2}\int
\frac{d({\kappa x^+})\kappa g'}{(\kappa g + {\kappa x^+} \kappa g') \mp
\sqrt{(\kappa g + {\kappa x^+} \kappa g')^{2} - 4M^2\kappa g'(\epsilon +
{\kappa x^+} {\kappa}g)/\bar{M}^2}},
\label{cored}
\end{equation}
where $\epsilon\approx e\Delta M/M$, and $\Delta M=\bar{M}-M$.

\medskip
\begin{center}
\leavevmode
\epsfxsize 2.5in
\epsfbox{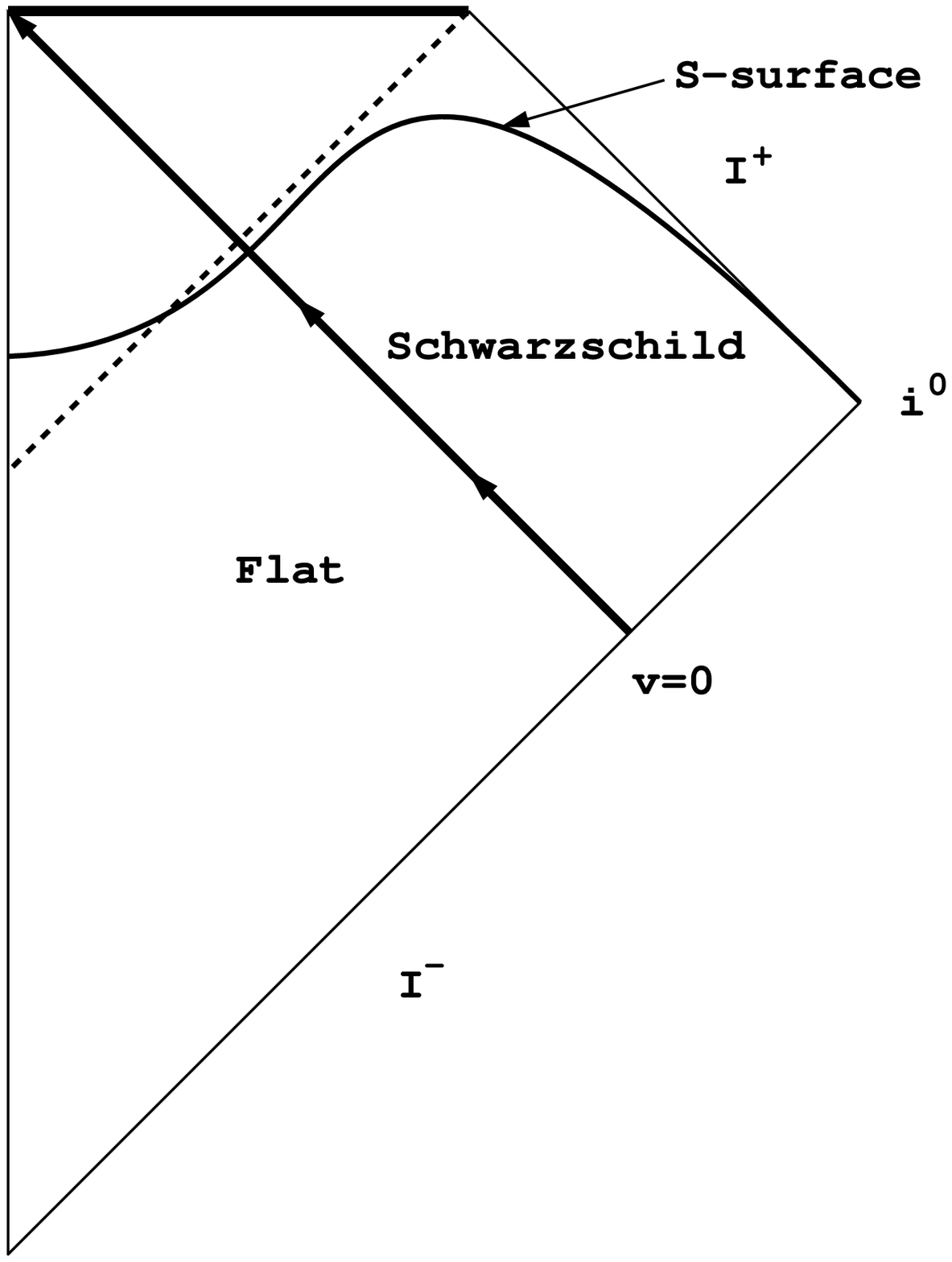}
\end{center}
\vskip 0.5 true cm
\begin{quotation}
\small
\noindent {\bf Figure 2:}
An S-surface in a Schwarzschild spacetime formed by a collapsing shell of
matter at $v=0$. The horizon is at $x^-=0$.
\end{quotation}
\medskip

As in the previous subsection, an S-surface can be specified by
\begin{equation}
\kappa g(\kappa x^+)=-\alpha^{2}\kappa x^{+} -\delta
\end{equation}
where the two constants $\alpha$ and $\delta$ are chosen so that $\alpha
\ll \delta \ll 1$, ensuring that the surface is almost lightlike, and runs
very close to the horizon. It follows that in the region of validity of our
approximation, $\kappa x^+\kappa x^-\lapproxeq e/M$ and $\kappa x^+>1$ in
which (\ref{cored}) is valid, $\kappa x^-$ effectively takes the constant
value $-\delta$.  Under the same condition, we can easily evaluate
(\ref{cored}) for $e/\Delta M\gapproxeq\kappa x^+>1$ so that
\begin{equation}
\ln(\bar{\kappa}\bar{x}^{+})=
\ln\left(\kappa x^{+} - \frac{\epsilon}
{\delta}\right) + C
\label{kcor}
\end{equation}
where $C$ is a constant of integration. Using $\kappa x^{+}=e^{\kappa
v}$, the above coordinate relationship translates to a coordinate
relationship between the null tortoise coordinates (that are asymptotically
flat on ${\cal{I}}^{+}$) of the form
\begin{equation}
\bar{\kappa}\bar{v}=
\ln\left(e^{\kappa v}-\frac{e\Delta M}{M\delta }\right) + C.
\label{asycore}
\end{equation}
This is the main result of this subsection, which we emphasize is valid on
the region of the S-surface above the collapsing shell and close to the
future horizon (see Fig. 2). A discussion of the shift for general black
hole spacetimes will be given in Ref. \cite{lo}.

\section{An Effective semiclassical Theory}

In this section we turn to the propagation of a matter state on a quantised
gravitational background. A state representing a black hole spacetime of
mass $M$ must, as discussed in the introduction, have an uncertainty in its
mass which is at the very least of the order of $1/M$ in Planck units (we
shall, for convenience, take it to be 1).  As discussed in \cite{lmo}, a
state for the gravitational field that is localised around a particular
classical black hole solution with mass $M_0$, can be regarded as a
superposition of plane wave WKB states, each of which is an approximate
mass eigenstate:
\begin{equation}
|\Psi_{\rm gravity}\rangle=\int d{M} \omega({M}) |{M}\rangle
\quad{\rm or}\quad
\Psi[h_{ij}]=\int d{M} \omega({M}) \Psi_{{M}}[h_{ij}]
\end{equation}
Here $\omega({M})$ is a weight function that is centered around a mean
value $M_0$ with a spread of order $1$.  We can approximate the propagation
of matter on a {\it quantum} black hole background by looking at
propagation on classical backgrounds with different masses in the
neighbourhood of $M_0$, so that the combined matter--gravity state is of
the form
\begin{equation}
|\Psi(\Sigma)\rangle = \int_{M_0-1}^{M_0+1} dM
\omega({M}) |{M}\rangle
|f_{{M}}(\Sigma)\rangle
\label{state}
\end{equation}
where $|f_{{M}}(\Sigma)\rangle$ is a Schr\"odinger picture state
propagating on a background spacetime ${\cal M}$ with mass ${M}$ \cite{banks}.

It is natural to trace over the gravitational degrees of freedom in
(\ref{state}) since fluctuations on the Planck scale are unobservable.
However, in normal semiclassical evolution, the propagation of matter
fields should be insensitive to such small fluctuations in the mass of the
background, so that all the $|f_{{M}}(\Sigma)\rangle$ for different $M$ are
approximately equal. In that case there is no entanglement between the
matter and gravity degrees of freedom and tracing leads to a pure state
density matrix, which is what we observe in everyday experiments. However,
we shall show that this is not always the case.

The simply seemingly simple statement of the previous paragraph that the
$|f_{{M}}(\Sigma)\rangle$ for different $M$ are approximately equal raises
the question of how we should compare matter states
$|f_{{M}}(\Sigma)\rangle$ propagating on different spacetimes with
different values of ${M}$. In canonical quantum gravity, a state
\begin{equation}
\Psi[h_{ij},f]
\end{equation}
takes the form of a correlation between a matter configuration and a
spacelike hypersurface ($h_{ij}$) without making reference to a {\it
spacetime}. Correlations between matter configurations and spacelike
hypersurfaces are also contained in the Schr\"odinger evolution of a
quantum matter field on any classical spacetime. These correlations should
be regarded as fundamental from the point of view of quantum
gravity. Through them a comparison of states defined on different
spacetimes, but on the same hypersurface (same $h_{ij}$) can be made. It is
for this reason that the embedding of a hypersurface in different
spacetimes was studied in the previous section.

A detailed discussion of the comparison of matter states defined on
neighbouring background spacetimes can be found in \cite{klmo} and
\cite{lmo}. The essential idea is the existence of a canonical Hilbert
space associated with a spacelike hypersurface, independent of any
spacetime. The value of a good old-fashioned Schr\"odinger picture state on
a particular hypersurface can be expressed as a state in this canonical
Hilbert space. Comparisons between Schr\"odinger picture states defined on
different spacetimes then also take place through this Hilbert space.

The Hilbert space associated with a surface $\Sigma$ is defined in terms of
a set of geometric modes on $\Sigma$ (for example of the form $e^{i\omega
s}$ where $s$ is the proper distance on $\Sigma_0$). At least in 2D, there
exists a canonical isomorphism between the Fock space defined with these
modes and the Hilbert space of all Schr\"odinger picture states, which
comes from evaluating the Schr\"odinger picture state on $\Sigma$ in an
appropriate foliation \cite{klmo}.

At past null infinity, we can consider a null hypersurface of constant $u$
or $x^-$, which we refer to as ${\cal I}^-$. Since everything is flat, the
Hilbert space intrinsic to ${\cal I}^-$ is isomorphic to the Hilbert space
of Schr\"odinger picture states evaluated at ${\cal I}^-$ in an obvious
way. Through the Hilbert space on ${\cal I}^-$ it is therefore
straightforward to say that states defined on different spacetimes are the
same at ${\cal I}^-$ -- they have the same Fock space decomposition with
respect to plane wave modes at past null infinity. When comparing states on
later-time hypersurfaces, we take the boundary condition that all states
are equal in this (intuitive and formal) sense at ${\cal I}^-$.

On later-time hypersurfaces, the comparison through the intrinsic Hilbert
space shows that states propagated on different spacetimes are no longer
identical. However, if the difference in mass between the spacetimes is
small, then for all practical purposes, the states are equal for almost any
choice $\Sigma$.  Under such circumstances, (\ref{state}) takes the form
\begin{equation}
|\Psi(\Sigma)\rangle\approx\left(\int
dM\;\omega(M)|M\rangle\right)|f_{M_0}(\Sigma)\rangle
\label{product}
\end{equation}
(where $M_0$ is a chosen mean value for $M$) and the semiclassical
approximation \cite{banks} of treating matter as propagating on a single
background spacetime is extremely good.

We can use (\ref{state}) to make sense of a trace over the unobserved
fluctuations in the gravitational background if we work in the
Schr\"odinger picture. For a state of the form (\ref{product}), tracing
over the $|M\rangle$ leaves a pure state density matrix
\begin{equation}
\rho_\Psi(\Sigma)=|f_{M_0}(\Sigma)\rangle\langle f_{M_0}(\Sigma)|
\label{pures}
\end{equation}
In general, however, tracing (\ref{state}) gives
\begin{equation}
\rho_\Psi(\Sigma) = \int d{M} \left\vert\,\omega({M})\right\vert^2
|f_M(\Sigma)\rangle\langle f_M(\Sigma)|
\label{dens}
\end{equation}
where one should think of all states $|f_M(\Sigma)\rangle$ as being in the
Hilbert space intrinsic to $\Sigma$.  The density matrix $\rho$ can be
considered the effective state of matter with the gravitational
fluctuations integrated out, at least when the departure from semiclassical
behaviour is not too large. Working with $\rho$ we can evaluate the
expectation value of operators in the usual way:
\begin{equation}
\langle {\cal O}\rangle = {\rm Tr}(\rho {\cal O})
\label{oop}
\end{equation}

For simplicity, let us assume that $\omega(M)$ is roughly constant over
some range $M_0-\Delta M_{max}\le M \ge M_0+\Delta M_{max}$ centered around
a mass $M_0$, with $\Delta M_{max}$ of order 1, and is zero outside this
range. The semiclassical approximation requires that states
$|f_M(\Sigma)\rangle$ of (\ref{state}) be approximately equal. We can test
this by evaluating the inner product between the elements
$|f_M(\Sigma)\rangle$ for a pair of values of $M$ within the specified
range. It was found in \cite{klmo} that this inner product is approximately
zero unless the difference between the values of $M$ differ at most on a
scale of the order of $e^{-M}$.

The breakdown in the semiclassical approximation implied by this result can
be expressed in terms of the entropy of the density matrix (\ref{dens}).
The entropy is zero for a pure state (\ref{pures}), but becomes non-zero as
the overlap
\begin{equation}
\langle f_M(\Sigma)|f_{\bar{M}}(\Sigma)\rangle
\label{overlap}
\end{equation}
between states in the density matrix decreases.  An estimate of the entropy
of this density matrix, assuming $\omega(M)$ to be of the form outlined
above, comes from looking at the number of approximately orthogonal states
that are contained in $\rho$. If for a given surface
\begin{equation}
\left\vert\langle f_M(\Sigma)|f_{\bar{M}}(\Sigma)\rangle
\right\vert^2
\ll 1
\label{overlap2}
\end{equation}
for a $\Delta M$ much smaller than 1 (the characteristic size of the
fluctuations in $M$), then we can approximate $\rho$ by changing the
integral in (\ref{dens}) to a sum over states that are approximately
orthogonal:
\begin{equation}
\rho_\Psi(\Sigma)={1\over N}\sum^{N/2}_{l=-N/2}
|f_{M_0+l\Delta M_0}(\Sigma)\rangle\langle f_{M_0+l\Delta M_0}(\Sigma)|
\end{equation}
where $\Delta M_0$ is the value of $M-M_0$ required for the overlap
(\ref{overlap}) to be effectively zero, and $N=\Delta M_{max}/\Delta M_0$,
where $\Delta M_{max}$ defines the size of the fluctuations in $M$ and is
of order 1.  This gives an estimate of the entropy ${\rm Tr}(\rho \ln
\rho)\sim \ln N$ of $\rho$. This entropy \cite{samir,klmo} can be
interpreted as the entanglement entropy between the gravitational and
matter degrees of freedom, and we refer to it as the gravitational entropy
associated with the surface $\Sigma$.

It is useful to define an operator that contains information about the
difference between states $|f_{{M}}(\Sigma)\rangle$ and
$|f_{\bar{M}}(\Sigma)\rangle$ on a hypersurface $\Sigma$. The simplest way
to define such an operator is to work in a Fock space picture. Any
Schr\"odinger picture state $|f_{{M}}(\Sigma)\rangle$ is associated with a
unique Fock state $|f_M\rangle$ on the Hilbert space of the spacetime with
mass $M$. Any state $|f_{\bar{M}}(\Sigma)\rangle$ can be associated with a
Fock state in the same Hilbert space in a hypersurface dependent way: For a
given $\Sigma$ define a Schr\"odinger picture state
$|g_{{M}}(\Sigma)\rangle$ such that it is equal to
$|f_{\bar{M}}(\Sigma)\rangle$ in the Hilbert space intrinsic to
$\Sigma$. Then associate $|f_{\bar{M}}(\Sigma)\rangle$ with the Fock state
$|g_{{M}}\rangle$ on the Hilbert space of the spacetime with mass $M$.

We can describe the difference between states $|f_{{M}}(\Sigma)\rangle$ and
$|f_{\bar{M}}(\Sigma)\rangle$ on a hypersurface $\Sigma$ by the operator
${\cal U}(\Sigma,\bar{M}\to M)$ so that
\begin{equation}
|f_{\bar{M}}\rangle \rightarrow |g_M\rangle \equiv {\cal
U}(\Sigma,\bar{M}\to M)|f_M\rangle
\end{equation}
Note that the operator is independent of the choice of state $|f_M\rangle$.
When $\bar{M}=M$, ${\cal U}$ is of course the identity operator for all
$\Sigma$. For $\bar{M}\ne M$, it implements the coordinate transformations
$\bar{x}^+(x^+)$ and $\bar{x}^-(x^-)$ \cite{klmo} and is constructed
explicitly in the next section. ${\cal U}$ defined in this way can be
thought of as an interaction picture time evolution operator (because of
its dependence on $\Sigma$) incorporating the quantum gravitational effects
on the matter states.

In \cite{klmo} the overlap (\ref{overlap}) was computed for a family of
S-surfaces. For these surfaces, ${\cal U}$ implements the shifts $x^+\to
a^+(x^++\Delta^+)$ and $x^-\to a^-(x^-+\Delta^-)$. The right-moving sector
of outgoing matter is in the Kruskal vacuum close to the horizon, and since
$\Delta^-$ is small for a small fluctuation, the shift $x^-\to
a^-(x^-+\Delta^-)$ is an approximate symmetry of the outgoing state. The
infalling matter, on the other hand, is in the Schwarzschild vacuum, and so
the transformation $x^+\to a^+(x^++\Delta^+)$ has a non-trivial effect that
becomes large as $\Delta^+$ grows. Thus for S-surfaces, for which
$\Delta^+$ is large, $\rho$ should represent a mixed state. For this reason
we shall focus exclusively on the infalling (or left-moving) matter, and we
shall take ${\cal U}$ to be the identity operator on the right-moving
sector.  Eq. (\ref{overlap2}) is satisfied for very small $\Delta M\ll1$
for S-surfaces \cite{klmo}, and this can be taken to indicate a breakdown
in the semiclassical approximation.

Note that since the large shift is confined to S-surfaces, the entanglement
between matter and gravity represented by ${\cal U}$ is a phenomenon
associated with specific foliations of spacetime \cite{klmo,lmo}. For
example, a foliation of the region close to the horizon that is associated
with a one-parameter family of $\alpha$'s and $\delta$'s of
Eq. (\ref{small}) that are not both small, does not give rise to a large
shift in the Kruskal coordinates, and ${\cal U}$ for such hypersurfaces is
trivial. The consequences of this are discussed later.

\section{The $\protect{\cal U}$ operator}

The effect of the operator ${\cal
U}(\Sigma,\bar{M}\to M)$ is to transform the vacuum state with respect to
modes $\phi_i(x^+)$ to the vacuum with respect to modes
$\phi_i(\bar{x}^+(x^+))$. Equivalently, the vacuum with respect to modes
$\phi_i(v)$ becomes that with respect to $\phi_i(\bar{v}(v))$ where as
before $\lambda v=\ln (\lambda x^+)$ and $\lambda\bar{v}=\ln (\lambda
\bar{x}^+)$.

Consider expanding the matter
field operator as either
\begin{equation}
\hat{\phi}(v)= \int_{0}^{\infty} \frac{d \omega}{\sqrt{2 \omega}}
 (a_{\omega} e^{-i\omega v}+ a^{\dagger}_{\omega} e^{i\omega v})
\label{aop}
\end{equation}
or
\begin{equation}
\hat{\phi}(v)= \int_{0}^{\infty} \frac{d \omega}{\sqrt{2 \omega}}
 (b_{\omega} e^{-i\omega\bar{v}(v)}+
 b^{\dagger}_{\omega} e^{i\omega\bar{v}(v)})
\label{bop}
\end{equation}
It follows that
\begin{equation}
b_{\omega}=
\int_{0}^{\infty} d \omega'(a_{\omega'} \alpha^{*}_{\omega' \omega}
+ a^{\dagger}_{\omega'}\beta^{*}_{\omega' \omega})
\label{bogrel}
\end{equation}
where $\alpha_{\omega \omega'}$ and $\beta_{\omega \omega'}$
 are the Bogoliubov coefficients:
\begin{equation}
\alpha_{\omega \omega'}=\frac{1}{2\pi}\sqrt{\frac{\omega'}{\omega}}
I^{+}_{\omega \omega'}\quad {\rm and} \quad
\beta_{\omega\omega'}=\frac{1}{2\pi}\sqrt{\frac{\omega'}{\omega}}
I^{-}_{\omega \omega'}
\end{equation}
and $I^{\pm}_{\omega \omega'}$  are the integrals \cite{klmo}
\begin{equation}
I^{\pm}_{\omega \omega'}=\int_{-\infty}^{\infty}
d\bar{v} e^{-i\omega v(\bar{v}) \pm i\omega' \bar{v}} .
\end{equation}
The vacuum defined by $a_{\omega}|0\rangle=0$ is then related to the state
${\cal U}(\Sigma,\bar{M}\to M)|0\rangle$ defined by
$b_{\omega}\;{\cal U}(\Sigma,\bar{M}\to M)|0\rangle=0$ by the
operator
\begin{equation}
{\cal U}(\Sigma,\bar{M}\to M)=B*\exp \left[-
\frac{1}{2}a^{\dagger}_{\omega'}D_{\omega' \omega} a^{\dagger}_{\omega}\right]
\label{uop1}
\end{equation}
where $D$ is the symmetric matrix
\begin{equation}
D_{\omega' \omega}=\int d\omega'' \beta^{*}_{\omega'' \omega'}
\alpha ^{-1}_{\omega \omega''}
\end{equation}
and $B$ is a normalization constant which is equal to $\langle 0|\,{\cal
U}(\Sigma,\bar{M}\to M)|0\rangle^{-1}$.

In order to compute the Bogoliubov coefficients $\alpha_{\omega \omega'}$
and $\beta_{\omega \omega'}$, the coordinate relation $\bar{v}(v)$ is
needed throughout $\Sigma$. However, working in a wave-packet basis the
relation $\bar{v}(v)$ in a restricted region is sufficient to approximate
the relevant components of the Bogoliubov coefficients. This allows one to
isolate local features of the operator ${\cal U}$ (in a wave packet basis
the index $\omega$ will be replaced by the indices $(j,n)$ \cite{giddnel}).
For $\lambda v > \Delta M/(\lambda\alpha{\scriptstyle \sqrt{M/\lambda}})$
and $\lambda v < (\alpha+{\scriptstyle
\sqrt{\bar{M}/\lambda}})/(\alpha+{\scriptstyle \sqrt{M/\lambda}})$, the
relevant $\beta$ coefficients are essentially zero. It follows that the
state $|0\rangle$ is only affected in a finite region, and that outside of
this region ${\cal U}$ is effectively the identity operator. For more
complicated spacetimes, such as the Schwarzschild solution, we cannot
obtain the exact relation across classical infalling matter, and even in
the CGHS case, it is not clear that we should trust our results entirely in
that region, but a local analysis still enables one to say a considerable
amount about ${\cal U}$.

One can see from (\ref{uop1}) that the operator ${\cal U}$ in a wave packet
basis creates `` localized particles '' in a region where the ${\beta}$
Bogoliubov coefficients are non zero. If we restrict our observation to a
certain region (say above the shock wave) then we lose the correlations
between the regions above and below the shock wave. For example, for
$\Delta M<0$, restricting attention to the region above the shock wave, but
below $\lambda v = \Delta M/(\lambda\alpha{\scriptstyle
\sqrt{M/\lambda}})$, the particle creation appears thermal. That is
\begin{equation}
\beta ^{*}_{jn\omega'} \approx -e^{-\pi \omega_{j}/\kappa} \alpha_{jn\omega'}.
\end{equation}
(here $n$ labels a wave-packet in this region). Note that the temperature
defined by this relation is the same as the temperature of the black
hole. The thermal form of ${\cal U}$ in this region comes about because of
a dramatic expansion in the map from the $\bar{v}$ to the ${v}$
coordinates. This expansion is compensated by an equally dramatic squashing
in the map in the region $0\ge \lambda v \ge\ln(\lambda x^+_<)$ since for
$v<\ln(\lambda x^+_<)$ the map is the identity. For $\Delta M>0$, the
reverse occurs, so that the operator ${\cal U}$ appears thermal for modes
in the interval $\ln(\lambda x^+_>) >v>0$ where there is an expansion in
the coordinate relation, but this is compensated by a region of squashing
for $v>\ln(\lambda x^+)$.  Similar results apply for the Schwarzschild
spacetime (with $\lambda$ replaced by $\kappa$) from the results of
Sec. 2.2. The use of the word thermal should not of course be taken to mean
that the operator ${\cal U}$ is not unitary.

There is another formal expression for
the operator ${\cal U}$ which is useful. For the region above the shock
wave it takes the form
\begin{equation}
{\cal U}(\Sigma,\bar{M}\to M)=\exp\left\{i\int dx^+ \Delta^+
T_{++}(x^+)\right\}
\label{uop}
\end{equation}
where $\Delta^+$ is the shift defined by (\ref{lshifts}) or
(\ref{gshifts}). From (\ref{uop})
 one can easily see that ${\cal U}$ is unitary.
It follows from Eqs. (\ref{aop}) and (\ref{bop}) and from
the commutation relation
\begin{equation}
\left[T_{++}(x^+_1),\phi(x^+_2)\right]=
i\delta(x^+_1-x^+_2)\partial_+\phi(x^+_2)
\end{equation}
that
\begin{equation}
b_i={\cal U}a_i{\cal U}^{-1},\qquad b^\dagger_i={\cal U}a^\dagger_i{\cal
U}^{-1}.
\end{equation}
and thus that $|0_{b}\rangle={\cal U}|0_{a}\rangle$ and that (\ref{uop1})
and (\ref{uop}) implement the same transformation.

We can now make contact with Ref. \cite{kvv}. There an analogous operator
${\cal U}$ is defined, which takes the form
\begin{equation}
{\cal U}_{KVV} = \exp\left\{i\int\int dx^+
dx^-T_{++}(x^+)T_{--}(x^-)\right\}
\label{kvv}
\end{equation}
The effect of this operator on the infalling matter is given by treating
$T_{--}$ as a classical perturbation on the background. With this
interpretation, (\ref{uop}) and (\ref{kvv}) are identical in the region
$x^+>1$ for $\Delta M<0$ (or $x^+>x^+_>$ for $\Delta M>0$). This can be
seen by noticing that $\Delta ^{+} \approx e^{\kappa u_{0}}$ where $u_{0}$
is the value of the retarded time characterising an S-surface. If we
take the energy-momentum tensor of the outgoing radiation in \cite{kvv} to
be $T_{uu}(u)=\Delta M \delta (u-u_{0})$, the correspondence becomes clear.

This correspondence arises despite the fact that the fluctuations in the
background are in our case due to fluctuations in the earlier infalling
matter and in Ref. \cite{kvv} due to fluctuations due to outgoing Hawking
radiation. The essential element of either calculation is that these
fluctuations are hugely amplified in their effect on the infalling state
close to the horizon\footnote{Note, however, that in our treatment we
regard ${\cal U}$ as having changed the {\it in} vacuum (the Schr\"odinger
picture), whereas in \cite{kvv} ${\cal U}$ changes the operators rather
than the states (the Heisenberg picture).}.

\section{Induced Energy-Momentum Tensor}

In this section we compute the expectation value of the energy-momentum
tensor of a state ${\cal U}(\Sigma,\bar{M}\to M)|0\rangle$ when
$\Sigma$ is an S-surface.  We have seen that on an S-surface the coordinate
relationship between the identified points on space times with mass $M$ and
mass $\bar{M}$ takes the form in the Kruskal coordinates
\begin{equation}
\bar{x} ^{+}= a^+(x^{+}+\Delta^+)
\end{equation}
and that the operator ${\cal U}$ implements a map between modes $\phi(v)$
and modes $\phi(\bar{v}(v))$. It follows from a standard calculation
that \cite{BnD}:
\begin{eqnarray}
\langle T_{vv}^{\bar{M}}(\Sigma) \rangle &\equiv& \langle 0|\;{\cal
U}^{-1}(\Sigma,\bar{M}\to M)\,T_{vv}\;{\cal U}(\Sigma,\bar{M}\to
M)|0\rangle
\\
&=&\frac{1}{12\pi}\left(\partial_v\bar{v}\right)^{1/2}
\partial_{v}^{2}\left(\partial_v\bar{v}\right)^{-1/2}
\label{tvv}
\end{eqnarray}
and all other components are zero.
Since $T_{vv}$ is conserved, this
quantity can be interpreted as the energy density of the state ${\cal
U}(\Sigma,\bar{M}\to M)|0\rangle$ at ${\cal I}^-$ if it were propagated
back freely on $M$, where it would correspond to a distribution of
infalling matter.

The simple expression (\ref{tvv}) allows us to compute the energy-momentum
tensor associated with the state ${\cal U}(\Sigma,\bar{M}\to M)|0\rangle$
in any local region with the knowledge of the coordinate transformation
$\bar{v}(v)$.  Assuming that $\bar{v}(v)$ has at least a continuous third
derivative one can integrate by parts so that
\begin{equation}
\int^{x_2}_{x_1} dv\langle T_{vv}^{\bar{M}}(\Sigma)  \rangle = \frac{1}{12}
\left.\left[\left(\partial_v\bar{v}\right)^{1/2}
\partial_{v}\left(\partial_v\bar{v}\right)^{-1/2}\right]\right|^{x_2}_{x_1} +
\frac{1}{48} \int^{x_2}_{x_1} dv
\left(\frac{\partial_v^2\bar{v}}{\partial_v\bar{v}}\right)^{2}
\end{equation}
Since the coordinate relationship is such that $\bar{v}(v) \longrightarrow
v$ when $v \longrightarrow \pm \infty$ (below the shock wave or at ${\cal
I}^+$), the total energy flux
\begin{equation}
E_{tot}^{\bar{M}}(\Sigma)  \equiv \int^{\infty}_{-\infty} dv \langle
T_{vv}^{\bar{M}} \rangle = \frac{1}{48\pi} \int^{\infty}_{-\infty}
\left(\frac{\partial_v^2\bar{v}}{\partial_v\bar{v}}\right)^2 > 0.
\label{etot}
\end{equation}
of the state must be positive.

Eqs. (\ref{lshifts}) and (\ref{gshifts}) give explicit expressions for
$\bar{v}(v)$ for the CGHS model when $\Sigma$ is an S-surface.  In the
three regions of (\ref{lshifts}) and (\ref{gshifts}), the function
$\bar{v}(v)$ has the form
\begin{equation}
\bar{v}(v)=\frac{1}{\lambda}\ln\left[e^{\lambda v}+\Delta^+\right]+{1\over
\lambda}\ln a.
\label{traj}
\end{equation}
However, because of the shock wave in the CGHS solution, the relation
$\bar{v}(v)$ is continuous but not differentiable.  With a smooth classical
infalling matter distribution, the function $\bar{v}(v)$ would be
smooth. Unfortunately it is much more difficult to obtain exact expressions
with a smoothened metric. Exact expressions can still be obtained when the
infalling matter arrives in a shock wave of some finite width instead of a
delta function, but this still leads to discontinuities in the
energy-momentum tensor of the same order of magnitude. We shall therefore
restrict our attention to the results obtained with the coordinate
relations (\ref{lshifts}) and (\ref{gshifts}). In this case we cannot
obtain explicit expressions for the total energy flux or for the
energy-momentum tensor in the neighbourhood of the singularities caused by
the shock wave.

Proceeding with (\ref{traj}) for $\bar{v}(v)$, one gets that
\begin{equation}
\langle T_{vv}^{\bar{M}} (\Sigma) \rangle = \frac{\lambda
^{2}}{48\pi}\left[1-\frac{1}{(1+ \Delta^+e^{-\lambda v})^{2}}\right].
\label{tvv1}
\end{equation}
in each of the three regions. From (\ref{lshifts}) and (\ref{gshifts}),
for $\Delta M <0$
\begin{equation}
\begin{array}{rcll}
\Delta^+&=&\displaystyle{1\over\alpha}\left(\sqrt{M\over\lambda}-\sqrt{\bar{M}\over\lambda}\right)
&\displaystyle\lambda v \ge 0
\\
\Delta^+&=&\displaystyle-\sqrt{\bar{M}\over\lambda}\left(\alpha+\sqrt{M/\lambda}\right)^{-1}
&\displaystyle0\ge\lambda v\ge \ln\left(\lambda x^+_<\right)
\\
\Delta^+&=&0
&\displaystyle\ln\left(\lambda x^+_<\right)\ge \lambda v
\end{array}
\label{cor1}
\end{equation}
where recall that $\lambda x^+_<=(\alpha
+{\scriptstyle\sqrt{\bar{M}/\lambda}})/(\alpha+
{\scriptstyle\sqrt{M/\lambda}})$. For $\Delta M>0$,
\begin{equation}
\begin{array}{rcll}
\Delta^+&=&\displaystyle-{1\over\alpha}\left(\sqrt{\bar{M}\over\lambda}-
\sqrt{{M}\over\lambda}\right)
&\displaystyle\lambda v\ge \ln\left(\lambda x^+_>\right)
\\
\Delta^+&=&\displaystyle{1\over\alpha}\sqrt{{M}\over\lambda}
&\displaystyle\ln\left(\lambda x^+_>\right)\ge\lambda v\ge 0
\\
\Delta^+&=&0
&\displaystyle 0\ge \lambda v
\end{array}
\label{cor2}
\end{equation}
where $\lambda x^+_> = 1+
({\scriptstyle\sqrt{\bar{M}/\lambda}-\sqrt{M/\lambda}})/\alpha$.
These expressions can be used to compute the value of $\langle
T_{vv}^{\bar{M}}(\Sigma)  \rangle$ in each region.
Recall that $\alpha$ is a very small number of the order of $e^{-M}$ in
Planck units.

Using (\ref{cor1}) and (\ref{cor2}), we see that for $\Delta M<0$, $\langle
T_{vv}^{\bar{M}}(\Sigma) \rangle$ is small at large $\lambda v$, but grows
as $\lambda v\to 0$. The region where $\langle T_{vv}^{\bar{M}} (\Sigma)
\rangle$ becomes non-negligible corresponds to the region $\lambda
v\lapproxeq \Delta M/(\lambda\alpha\scriptstyle\sqrt{M/\lambda})$ where the
$\beta$ coefficients become non-zero and ${\cal U}$ starts to appear
thermal in a wave-packet basis. In this region we can derive a similar
result for S-wave propagation on a Schwarzschild spacetime using
(\ref{asycore}). Throughout this region, down to $\lambda v=0$, $\langle
T_{vv}^{\bar{M}}(\Sigma) \rangle\sim\lambda^2/48\pi$. In the small region
$0>\lambda v>\ln(\lambda x^+_<)$, $\langle T_{vv}^{\bar{M}}(\Sigma)
\rangle$ is negative and very large, taking the value $-\lambda
M/48\pi\alpha^2$ at $\lambda v=\ln(\lambda x^+_<)$. Summing the
contributions from these two regions, the total energy flux on the surface
appears to be negative, but (\ref{etot}) implies that the singular points
must contribute a large energy flux to compensate for the total negative
energy flux of (\ref{tvv1}).

For the case $\Delta M>0$, $\langle T_{vv}^{\bar{M}}(\Sigma) \rangle$ is
again zero for large $\lambda v$, but becomes large and negative as
$\lambda v$ enters the region where the coordinate transformation
$\bar{v}(v)$ is non-trivial. The fact that in this case the energy flux is
negative can be understood because the transformation $\bar{v}(v)$ squashes
a large interval in $\bar{v}$ to a small interval in ${v}$. For $\lambda
v=\ln(\lambda x^+_>)$, $\langle T_{vv}^{\bar{M}}(\Sigma)
\rangle\sim-\lambda^2\Delta M^2/192\pi \alpha^2 M$. In the region
$\ln(\lambda x^+_>)>\lambda v>0$ there is a positive energy flux of order
$\lambda^2/48\pi$. The total energy flux is negative, so that again the
singular points must contribute a large energy flux . Of course as $\Delta
M\to 0$ from above or below the energy-momentum tensor in each region
approaches zero.

\medskip
\begin{center}
\leavevmode
\epsfxsize 6in
\epsfbox{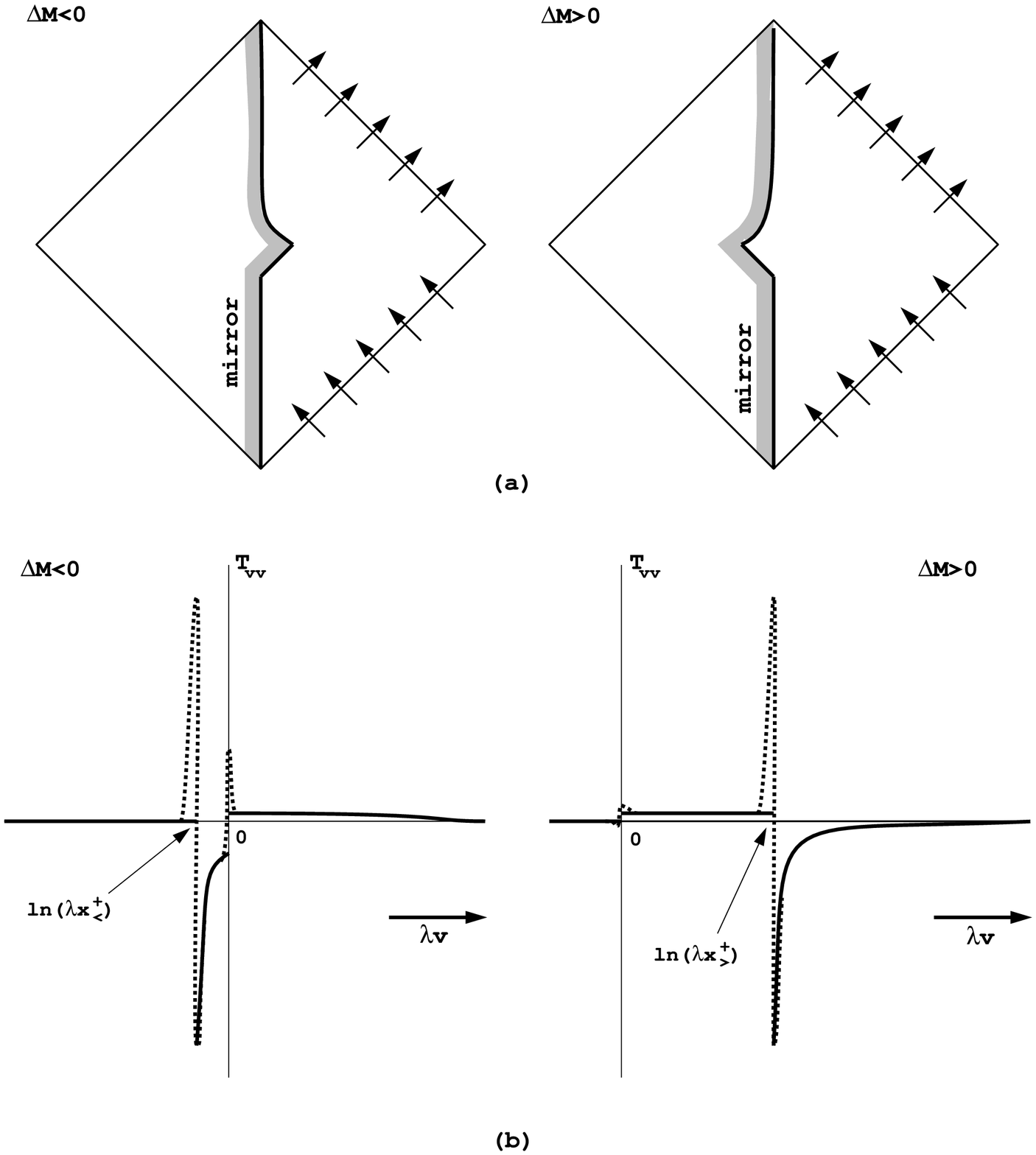}
\end{center}
\vskip 0.5 true cm
\begin{quotation}
\small
\noindent {\bf Figure 3:} (a) The mirror trajectories. (b) The energy flux
distribution for the cases $\Delta M<0$ and $\Delta M>0$. The undotted
lines represent (\ref{tvv1}). The dotted lines are an extrapolation in the
vicinity of the jumps in the coordinate relations (\ref{lshifts}) and
(\ref{gshifts}) where some energy flux must be present according to
(\ref{etots}).
\end{quotation}
\medskip

Eq. (\ref{tvv}) for $\langle T_{vv}^{\bar{M}}(\Sigma) \rangle$ is most
familiar from the problem of quantum field theory in the presence of a
moving mirror. It is useful to translate the coordinate transformations
$\bar{v}(v)$ given by (\ref{lshifts}) and (\ref{gshifts}) into the moving
mirror language to get an intuitive picture of the generation of energy by
the operator ${\cal U}$.

Given a continuous function $\bar{v}(v)$ we can find a mirror trajectory
that converts the initial vacuum (at ${\cal{I}}^{-}$) to the state ${\cal
U}(\Sigma,\bar{M}\to M)|0\rangle$ (at ${\cal{I}}^{+}$).  The trajectory
$x=z(t)$ satisfies
\begin{equation}
\bar{v}(v)-v=2z\left(\frac{v+\bar{v}(v)}{2}\right)
\end{equation}
One can solve for the trajectory of the mirror. For the general
$\bar{v}(v)$ given by (\ref{traj}) one finds that
\begin{equation}
2\lambda z(t)=\ln \left\{ \frac{\Delta^++\sqrt{(\Delta^+)^2 +4e^{2\lambda
t}/a}}
{-\Delta^++\sqrt{(\Delta^+)^2 +4e^{2\lambda t}/a}} \right\} +\ln a
\label{mirror}
\end{equation}
In Figure 3 we give a plot of the mirror trajectory and of $\langle
T_{vv}\rangle$ associated with each of the trajectories. It is clear from
the figure that the mirror trajectory is not smooth. The kinks in the
trajectory must contribute enough energy flux to surpass the lower bound
given by (\ref{etots}).

We can use (\ref{etot}) itself to estimate the positive energy flux
that comes from the singular regions. For the coordinate relationship
(\ref{traj}),
\begin{equation}
\partial_v^2\bar{v}=\lambda\partial_v\bar{v}\left(1-\partial_v\bar{v}\right)
\end{equation}
Thus a lower bound on the total energy flux is given by
\begin{equation}
E^{\bar{M}}_{tot}(\Sigma)={\lambda^2\over
48\pi}\left[\int_{-\infty}^{v_1}\left(1-\partial_v\bar{v}\right)^2
+\int_{v_1}^{v_2}\left(1-\partial_v\bar{v}\right)^2
+\int_{v_2}^{\infty}\left(1-\partial_v\bar{v}\right)^2\right]
\label{etots}
\end{equation}
where $v_1$ and $v_2$ are the boundaries of the regions for which
$\bar{v}(v)$ is smooth. This gives a lower bound of
\begin{equation}
E_{tot}^{\bar{M}}(\Sigma)\gapproxeq\frac{\lambda^2}{96 \pi} \frac{2|\Delta
M/\lambda|\sqrt{M/\lambda}}{\alpha(2\alpha\sqrt{M/\lambda}+|\Delta M/\lambda|)}
\label{energ2}
\end{equation}
for $\bar{M}<M$, and of
\begin{equation}
E_{tot}^{\bar{M}}(\Sigma)\gapproxeq \frac{\lambda^2}{48 \pi}\frac{\Delta
M/\lambda}{2\alpha \sqrt{M/\lambda}}
\label{energ1}
\end{equation}
for $\bar{M}>M$.

Evaluating the ensemble average of the total energy flux (in the sense of
equation (\ref{oop})) we find that
\begin{equation}
<E_{tot}(\Sigma)> \sim \frac{1}{\alpha \sqrt{M/\lambda}}= e^{\lambda u_{0}}
\end{equation}
where $\lambda u$ is the retarded time at ${\cal I}^{+}$ and $\lambda
u_{0}$ is a particular value depending on the S-surface.  It is the value
of $\lambda u$ where the S-surface intersects the shock wave and, because
of the nature of an S-surface, throughout the region where there is a large
induced energy-momentum tensor. Now if the total induced energy flux
becomes of the order of the mass of the black hole we would expect that the
semiclassical approximation ceases to be valid. In particular we expect the
Hawking radiation to be affected. This gives a characteristic time for the
departure of the Hawking radiation from its expected form of
\begin{equation}
u_{0} \sim \frac{1}{\lambda} \ln (M/\lambda)
\end{equation}
for the CGHS black hole. For the Schwarzschild black hole, using the
results of Sec. 2.2, we can estimate this time to be
\begin{equation}
u_{0}^{Sch} \sim M \ln M
\end{equation}
(remember that $M$ is the black hole mass in Planck units).

The fact that the expectation value of $T_{vv}$ becomes large suggests that
outside observers should see very different spacetime dynamics to what one
would expect. This might be connected to the `bounce' model discussed in
\cite{hws}.

It is unclear how seriously we should take the precise form of $\langle
T_{vv}^{\bar{M}}(\Sigma) \rangle$ that we have computed here. The fact that
the absolute value of $\langle T_{vv}^{\bar{M}}(\Sigma) \rangle$ is
extremely large for different $M$ and sensitive to $\Delta M$ indicates
that the semiclassical approximation is breaking down in a dramatic way,
since the different states ${\cal U}(\Sigma,M\to M_0)|0\rangle$ in
(\ref{dens}) contribute to an expectation value for the operator $T_{vv}$
(in the sense of (\ref{oop})) that has huge quantum fluctuations. The
expectation value for the total energy flux on an S-surface (\ref{energ2})
and (\ref{energ1}) are clearly so large that the methods used to compute
them are no longer consistent.  Nevertheless, for both $\Delta M<0$ and
$\Delta M>0$ we can have confidence in our results in the region of large
$\lambda v$ far from the shock wave where, as we have seen, $\langle
T_{vv}^{\bar{M}}(\Sigma) \rangle$ starts to become large. The results we
obtain there are robust and can be rederived for Schwarzschild spacetime
using the results of the Sec. 2.2.

The large fluctuations in the energy density of the infalling matter
associated with S-surfaces can drastically affect the background metric as
seen by an outside observer. There appears to that observer to be a
significant interaction between incoming matter and outgoing Hawking
radiation, and this interaction may make the evolution of matter fields
appear unitary. On the other hand, for generic foliations in the vicinity
of the horizon there is no breakdown in the semiclassical approximation,
and no large backreaction, suggesting that an infalling observer falls
through the horizon according to semiclassical expectations. The picture
that emerges from these very different predictions about the physics of the
event horizon is remarkably similar to the ideas underlying the principle
of black hole complementarity \cite{thooft,stu,kvv,hws}. The notion of
complementarity appears to arise naturally from considering the gravity
sector to be quantized and in that sense is not tied to any specific
formulation of quantum gravity.

{}From the perspective of an outside observer, the large fluctuations in
energy close to the horizon can be used to delineate the boundary between
the region where semiclassical physics is valid, and the region
where quantum gravity effects have a dramatic (although
largely uncomputable) effect. The location of this boundary, and its
appearance to outside observers, is the subject of the next section.

\section{Stretched Horizon}

It is instructive to evaluate the total energy-momentum tensor in the sense
of (\ref{oop}) for large $\lambda v$:
\begin{equation}
\langle{T}_{vv}(\Sigma)\rangle=-\frac{\lambda ^{2}}{48 \pi} \left[\frac{1}
{16(M/\lambda) \alpha ^{2} e^{2\lambda v} -1}\right]
\label{enerabov}
\end{equation}
If on an S-surface we go above a region obeying
\begin{equation}
\lambda x^{+}=e^{\lambda v}=\frac{1}{2\alpha\sqrt{M/\lambda}}
\end{equation}
the induced energy-momentum tensor is small. This region also coincides
with the region where the $\beta$ coefficients are almost zero.  It follows
that on each S-surface labeled by $\alpha$ there is an approximate value of
$\lambda v$ below which there is a problem with the semiclassical
approximation (and the concept of a background spacetime is probably
absent), but above which there are no local quantum gravity effects. These
points, one for each S-surface, define a {\it timelike} hypersurface
marking the boundary between a semiclassical region and a region of strong
quantum gravitational effects. Since the approximate value of $\lambda
x^{-}-M/\lambda$ is $-2\alpha\sqrt{M/\lambda}$ for an S-surface defined by
$\alpha$ in the vicinity of the horizon, the boundary surface obeys the
equation
\begin{equation}
-\lambda x^{+} (\lambda x^{-}-M/\lambda )\approx 1
\label{stho}
\end{equation}
If we interpret $e^{2\phi}$ to be related to the area of a spherical
hypersurface in a four dimensional space time, then by (\ref{cghs}) this
boundary surface is a constant area hypersurface of one planck unit of area
greater than the event horizon (see Fig. 4).

We can estimate the location of the boundary surface between semiclassical
and quantum gravitational regions for the Schwarzschild black hole, as seen
by an outside observer, using the expression (\ref{asycore}) for coordinate
shift derived in Sec. 2.2.  From equation (\ref{kcor}) we see that there is
little change in the state for points on an S-surface (now labeled by
$\delta$), obeying $\kappa x^{+} >{4M\Delta M}/{\delta}$ (remember that the
analogue of $\delta$ in the CGHS is $\alpha
\scriptstyle\sqrt{M/\lambda}$). As the value of $\kappa x^{-}$ for the
S-surface throughout this region is approximately $\delta$, we find that
the boundary surface is a timelike hypersurface obeying
\begin{equation}
-\kappa x^{+} \kappa x^{-} \approx 4Me \Delta
M_{max} \approx 2eM.
\end{equation}
It follows from equation (\ref{met}) that this is a constant $r$
hypersurface approximately one Planck length away from the horizon. The
distance between the event horizon and this `stretched horizon', in the
language of Ref.  \cite{thorne}, is determined by the magnitude of the
fluctuations of the variables in the gravitational sector (in this case the
black hole mass).

For a computation that involves quantities outside the boundary surface,
for instance some S matrix elements between states at ${\cal{I}^{-}}$ and
states at ${\cal{I}^{+}}$ we can attempt to use the semiclassical
approximation, as long as we are not interested in the regions which are to
the future of the boundary surface.  However, almost all of the Hawking
radiation is detected to the future of the boundary surface.

\medskip
\begin{center}
\leavevmode
\epsfxsize 4in
\epsfbox{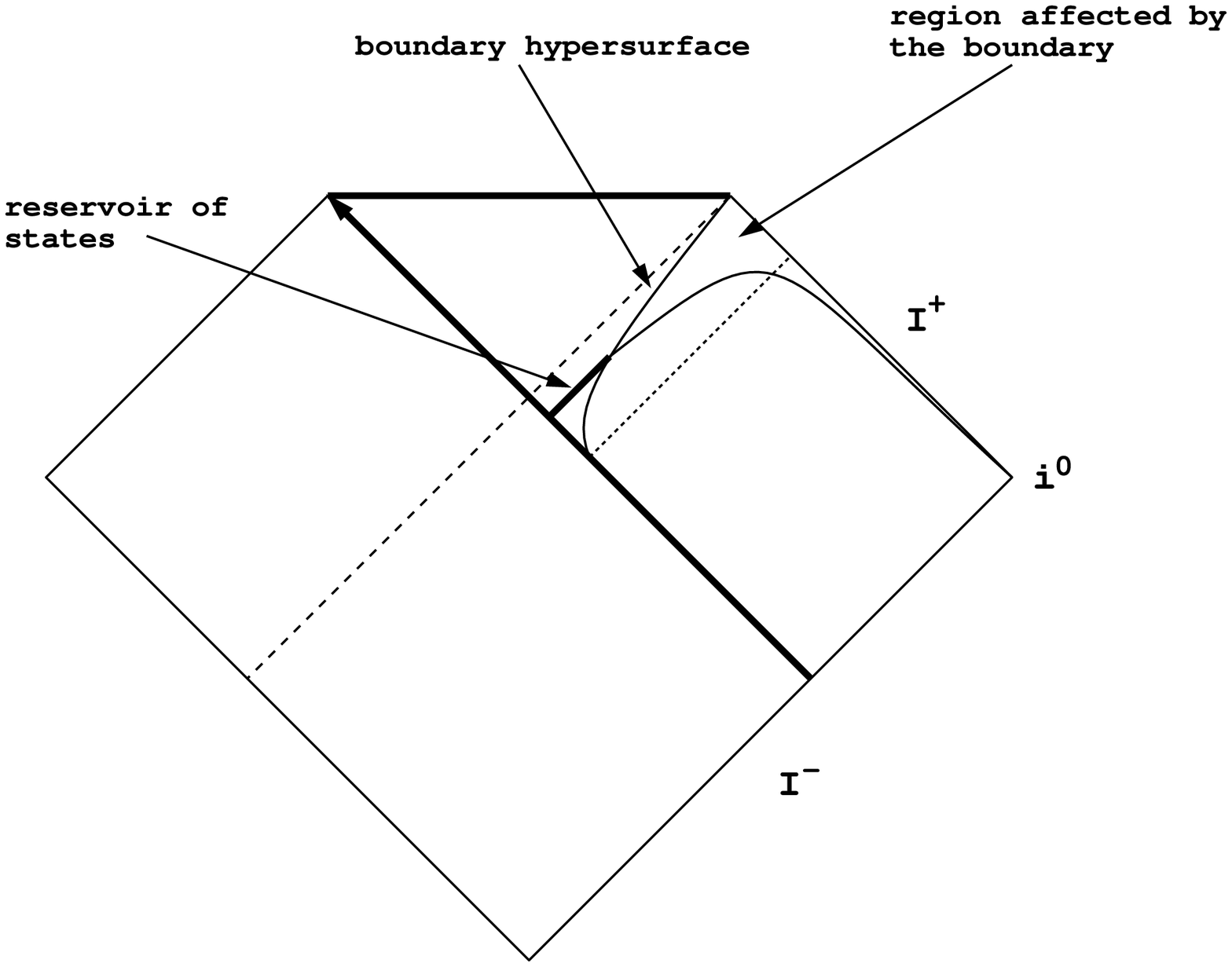}
\end{center}
\vskip 0.5 true cm
\begin{quotation}
\small
\noindent {\bf Figure 4:}
The boundary curve or stretched horizon in shown for the CGHS spacetime.
\end{quotation}
\medskip

In order to make any predictions to the future of the boundary surface, we
need to use quantum gravity. At this point, there are a couple of
possibilities. On the one hand, it may be that the full details of the
quantum gravitational interactions behind the stretched horizon are
necessary for any calculations to its future, and that no conclusions can
be reached about black hole evaporation without a theory that models these
interactions. On the other hand, it may be, as has been speculated recently
\cite{stu}, that a theory defined {\it on} the stretched horizon is
sufficient to model all observations made outside the horizon.

Whichever of these scenarios is correct, it is interesting to consider as
simple an approach as possible to modeling the physics of the stretched
horizon. We should try to work on a single background spacetime, but then
we must trace over the microstates of the gravitational field. The
in-falling matter states should be treated as being in a statistical
ensemble ({\it i.e.}  $\rho$), so that any outgoing state will interact
with this reservoir of states in a very complicated way. As we do whenever
a system with a large numbers of degrees of freedom is coupled to a smaller
system, we can attempt invoke an effective description of the microstates,
using statistical mechanics or thermodynamics. Usually a thermodynamical
description is valid if the interaction between the microstates and the
small system is weak and if the number of microstates is large enough.  In
the case of the black hole the second criterion certainly holds as from our
calculation one can see that the number of different states at the horizon
(the microstates, now thought of as residing in the in-falling matter
beyond the boundary surface) is huge. One can then hope to be able to apply
statistical mechanics and describe the effective interaction between the
gravity fluctuations (that manifest themselves in the form of a large
number of horizon states) and matter fields propagating on the black hole
background by ascribing energy, entropy and/or temperature to the boundary
surface. It is plausible that this may be the origin of the first law of
black hole thermodynamics, and is similar to the stretched horizon idea
advocated in \cite{stu}.  In \cite{lo} we shall argue that this procedure
associates a temperature to the stretched horizon, and that the log of
number of different horizon states should be associated with the Bekenstein
entropy of the black hole.

\section{Conclusions}

We have derived a series of results related to the computation of the state
of matter on certain hypersurfaces in a black hole spacetime. These
hypersurfaces are characterized by the fact that they yield simultaneous
information about the state of matter at future null infinity and close to
the event horizon, and are called S-surfaces.

We analyzed the breakdown of the semiclassical approximation for the CGHS
model (and for the S-wave matter sector on a Schwarzschild background) on
S-surfaces. It was shown in \cite{klmo} that small fluctuations in the mass
of a black hole lead to an entanglement between gravitational and matter
degrees of freedom. We have shown here that tracing over the gravitational
fluctuations leads to an effective semiclassical description in which a
matter state is described by a density matrix rather than a pure
state. Each element in the density matrix can be interpreted as the result
of a transformation of the original matter state by an operator ${\cal
U}$. This operator was constructed explicitly and is closely related to the
operator ${\cal U}$ defined in Ref. \cite{kvv} to represent the effect of
Planckian interactions near a black hole horizon. We have also shown that
there is a large induced energy-momentum tensor associated with the
transformed states. This result confirms that there is a dramatic breakdown
in the semiclassical picture on S-surfaces, and shows that quantum gravity
effects can lead to a large back-reaction close to the horizon. Finally, we
identified the region where quantum gravity effects are important, and
argued that the boundary of that region is naturally identified as a
stretched horizon. Some work related to the results given in this paper can
be found in Ref. \cite{km}.

The results described are valid only for S-surfaces. A generic foliation of
the region close to the horizon would not give rise to a breakdown in the
semiclassical approximation. In this sense, the state of matter is not
covariant in the usual way, and the complementarity idea arises naturally
(see \cite{lmo} for further discussion of this point). The stretched
horizon only has physical meaning for observers that do not cross the
horizon.

The quantum gravitational interactions to which we have referred involve
degrees of freedom associated with the state of the gravitational
background. It is important to note that the quantum nature of the
background is closely tied to that of the matter state that originally
formed the black hole, since the latter is largely responsible for what we
have termed gravitational fluctuations. Thus we can think of there being an
entanglement between the degrees of freedom of the matter that created the
black hole and the quantum fields that propagate on the hole. The
entanglement gives rise to a reservoir of states that can significantly
affect the Hawking radiation. In this paper we have only considered the
degree of freedom associated with the energy of the matter that forms the
black hole (contained in our discussion in the function $\omega(M)$ of
(\ref{state})). More generally one should consider the effect of
fluctuations in all the degrees of freedom of the matter forming the black
hole on in- and out-going quantum fields.

Finally, it is suggested that an outside observer might describe the
complex interactions behind the stretched horizon by an effective theory in
which matter interacts with a stretched horizon endowed with
thermodynamical variables. An effective picture of this kind ignores the
details of the microstates behind the stretched horizon so could not lead
directly to a unitary picture of black hole evaporation\footnote{One might
instead consider assigning the degrees of freedom of the entire reservoir
of states to the stretched horizon, perhaps leading to a unitary theory.},
but may give a microstate explanation for the laws of black hole
thermodynamics.

\section*{Acknowledgements}
We thank Larry Ford, Esko Keski-Vakkuri and Samir Mathur for helpful
discussions.

\end{document}